\documentclass[10pt,journal,compsoc]{IEEEtran}
\ifCLASSOPTIONcompsoc
  \usepackage[nocompress]{cite}
\else
  \usepackage{cite}
\fi
\ifCLASSINFOpdf
\else
\fi
\usepackage[mathscr]{euscript}
\usepackage{wrapfig}
\usepackage{color, colortbl}
\definecolor{lightgray}{gray}{0.9}
\usepackage[]{algorithm2e}
\usepackage{algorithmicx}
\usepackage{algpseudocode}
\usepackage{xcolor}
\usepackage{amsfonts}
\usepackage{amssymb}
\usepackage{amsmath}
\usepackage{mathptmx}
\usepackage{amsthm}
\usepackage{graphicx}
\usepackage{hyperref}

\hypersetup{
    colorlinks=true,
    citecolor=cyan,
    linkcolor=magenta,
    filecolor=magenta,
    urlcolor=cyan,
}
\hypersetup{draft}
\let\oldnl\nl
\newcommand{\nonl}{\renewcommand{\nl}{\let\nl\oldnl}}
\newcommand{\figref}[1]{Figure~\ref{#1}}
\newcommand{\tabref}[1]{Table~\ref{#1}}
\newcommand{\eqnref}[1]{Eq.~\eqref{#1}}
\newcommand{\secref}[1]{Section~\ref{#1}}

\newcommand{\lstref}[1]{Algorithm~\ref{#1}}

\newcommand{\code}[1]{\textbf{\texttt{#1}}}

\newcommand{\comments}[1]{}
\newlength\savedwidth
\newcommand\whline[1]{\noalign{\global\savedwidth\arrayrulewidth
                               \global\arrayrulewidth #1} %
                      \hline
                      \noalign{\global\arrayrulewidth\savedwidth}}

\begin{document}
%
\title{H-CNN: Spatial Hashing Based CNN for 3D Shape Analysis}
%
%
%
%

\author{Tianjia~Shao,
        Yin~Yang,
        Yanlin~Weng,
        Qiming~Hou,
        Kun~Zhou
        }
\IEEEtitleabstractindextext{%
\begin{abstract}
We present a novel spatial hashing based data structure to facilitate 3D shape analysis using convolutional neural networks (CNNs). Our method well utilizes the sparse occupancy of 3D shape boundary and builds hierarchical hash tables for an input model under different resolutions. Based on this data structure, we design two efficient GPU algorithms namely \code{hash2col} and \code{col2hash} so that the CNN operations like convolution and pooling can be efficiently parallelized. The spatial hashing is nearly minimal, and our data structure is almost of the same size as the raw input. Compared with state-of-the-art octree-based methods, our data structure significantly reduces the memory footprint during the CNN training. As the input geometry features are more compactly packed, CNN operations also run faster with our data structure. The experiment shows that, under the same network structure, our method yields comparable or better benchmarks compared to the state-of-the-art while it has only one-third memory consumption. Such superior memory performance allows the CNN to handle high-resolution shape analysis.
\end{abstract}

}

\maketitle

\IEEEdisplaynontitleabstractindextext

%
\IEEEpeerreviewmaketitle


\section{Introduction}
\label{sec:intro}
3D shape analysis such as classification, segmentation, and retrieval has long stood as one of the most fundamental tasks for computer graphics. While many algorithms have been proposed (e.g. see~\cite{hu2016siggraph}), they are often crafted for a sub-category of shapes by manually extracting case-specific features. A general-purpose shape analysis that handles a wide variety of 3D geometries is still considered challenging.
%
%
%
On the other hand, convolutional neural networks (CNNs) are skilled at learning essential features out of the raw training data. They have demonstrated great success in many computer vision problems for 2D images/videos~\cite{krizhevsky2012imagenet,simonyan2014very,long2015fully}. The impressive results from these works drive many follow-up investigations of leveraging various CNNs to tackle more challenging tasks in 3D shape analysis.

%

Projecting a 3D model into multiple 2D views is a straightforward idea which maximizes the re-usability of existing 2D CNNs frameworks~\cite{bai2016gift,qi2016volumetric,su2015multi,shi2015deeppano}. If the input 3D model has complex geometry however, degenerating it to multiple 2D projections could miss original shape features and lower quality of the final result. It is known that most useful geometry information only resides at the surface of a 3D model. While embedded in $\mathbb{R}^3$, this is essentially two-dimensional. Inspired by this fact, some prior works try to directly extract features out of the model's surface~\cite{masci2015geodesic,boscaini2015learning} using, for instance the Laplace-Beltrami operator~\cite{vallet2008spectral}. These methods assume that the model's surface be second-order differentiable, which may not be the case in practice. In fact, many scanned or man-made 3D models are of multiple components, which are not even manifolds with the presence of a large number of holes, dangling vertices and intersecting/interpenetrating polygons. Using dense voxel-based discretization is another alternative~\cite{wu20153d,maturana2015voxnet}. Unfortunately, treating a 3D model as a voxelized volume does not scale up as both memory usage and computational costs increase cubically with the escalated voxel resolution. The input data would easily exceed the GPU memory limit under moderate resolutions.

%

Octree-based model discretization significantly relieves the memory burden for 3D shape analysis~\cite{riegler2017octnet,OCNN_2017}. For instance, Wang et al.~\cite{OCNN_2017} proposed a framework named O-CNN (abbreviated as OCNN in this paper), which utilizes the octree to discretize the surface of a 3D shape. In octree-based methods, whether or not an octant is generated depends on whether or not its parent octant intersects with the input model. As a result, although octree effectively reduces the memory footprint compared to the ``brute-force'' voxelization scheme, its memory overhead is still considerable since many redundant empty leaf octants are also generated, especially for high-resolution models.

\begin{figure*}[t!]
\centering
  \includegraphics[width=\textwidth]{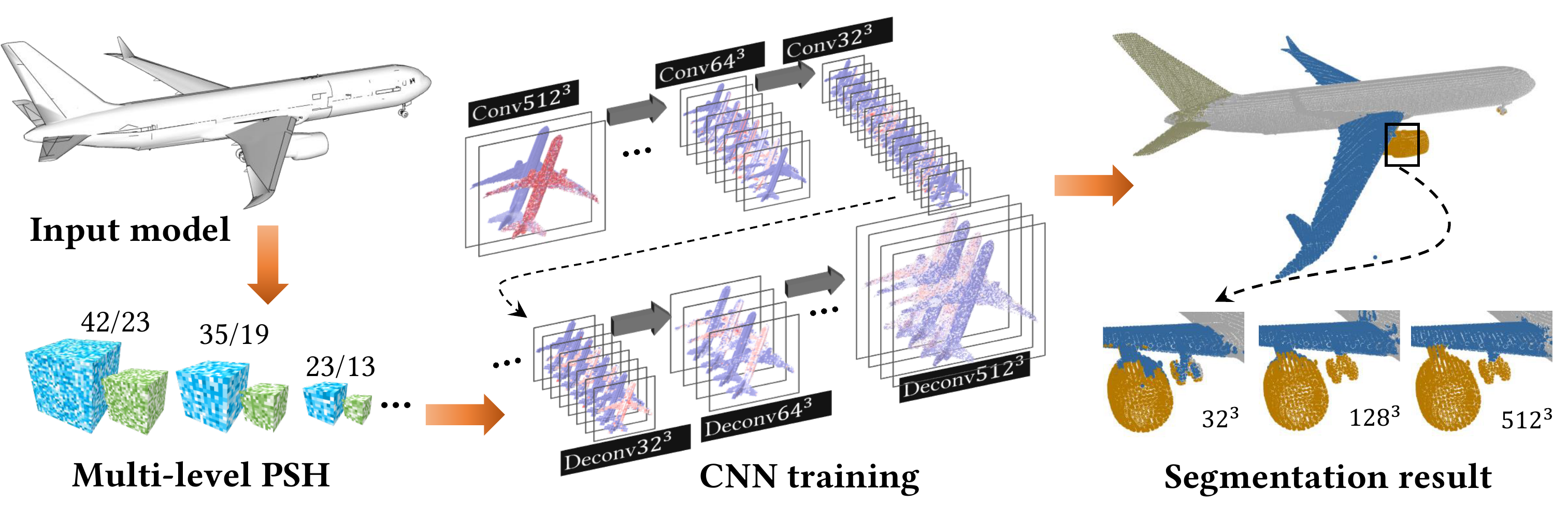}
  \caption{An overview of HCNN framework for shape analysis. We construct a set of hierarchical PSHs to pack surface geometric features of an input airplane model at different resolution levels. Compared with existing 3D CNN frameworks, our method fully utilizes the spatial sparsity of 3D models, and the PSH data structure is almost of the same size as the raw input. Therefore, we can perform high-resolution shape analysis with 3D CNN efficiently. The final segmentation results demonstrate a clear advantage of high-resolution models. Each part of the airplane model is much better segmented at the resolution of $512^3$, which is currently only possible with HCNN.}
  \label{fig:teaser}
\end{figure*}

In this paper, we provide a better answer to the question of how to wisely exploit the sparse occupancy of 3D models and structure them in a way that  conveniently interfaces with various CNN architectures, as shown in~\figref{fig:teaser}. In our framework, 3D shapes are packed using the perfect spatial hashing (PSH)~\cite{PSH_2006} and we name our framework as Hash-CNN or HCNN. PSH is \emph{nearly minimal} meaning the size of the hash table is almost the same as the size of the input 3D model. As later discussed in \secref{subsec:memory_analysis}, our memory overhead is tightly bounded by $\mathbf{O}(N^{\frac{4}{3}})$ in the worst case while OCNN has a memory overhead of $\mathbf{O}(N^2)$, not to mention other $\mathbf{O}(N^3)$ voxel-based 3D CNNs (here, $N$ denotes the voxel resolution at the finest level). Due to the superior memory performance, HCNN is able to handle high-resolution shapes, which are hardly possible for the state-of-the-art. Our primary contribution is investigating how to efficiently parallelize CNN operations using hash-based models. To this end, two GPU algorithms namely \code{hash2col} and \code{col2hash} are contrived to facilitate CNN operations like convolution and pooling.
Our experiments show that HCNN achieves comparable or better benchmarks under various shape analysis tasks compared with existing 3D CNN methods. In addition, HCNN consumes much less memory and it also runs faster due to its compact data packing.

\section{Related Work}
\label{sec:related_work}

3D shape analysis~\cite{Mitra2014ShapeAnal,Xu2016ShapeAnal,hu2016siggraph} is one of the most fundamental tasks in computer graphics. Most existing works utilize manually crafted features for dedicated tasks such as shape retrieval and segmentation. Encouraged by great successes in 2D images analysis using CNN-based machine learning methods~\cite{sharif2014cnn,he2015delving,girshick2016region}, many research efforts have been devoted to leverage CNN techniques for 3D shape analysis.


A straightforward idea is to feed multiple projections of a 3D model as the CNN input~\cite{bai2016gift,qi2016volumetric,su2015multi,shi2015deeppano} so that the existing CNN architectures for 2D images can be re-used. However, self-occlusion is almost inevitable for complicated shapes during the projection, and the problem of how to faithfully restore complete 3D information out of 2D projections remains an unknown one to us.

Another direction is to perform CNN operations over the geometric features defined on 3D model surfaces~\cite{bronstein2017geometric}. For instance, Boscaini et al.~\cite{boscaini2015learning} used windowed Fourier transform and Masci et al.~\cite{masci2015geodesic} used local geodesic polar coordinates to extract local shape descriptors for the CNN training. These methods, however require that input models should be smooth and manifold, and therefore cannot be directly used for 3D models
composed of point clouds or polygon soups. Alternatively, Sinha et al.~\cite{sinha2016deep} parameterized a 3D shape over a spherical domain and re-represented the input model using a geometry image~\cite{gu2002geometry}, based on which the CNN training was carried out. Guo et al.~\cite{guo20153d} computed a collection of shape features and re-shaped them into a matrix as the CNN input. Recently, Qi et al.~\cite{pointnet} used the raw point clouds as the network input, which is also referred to as PointNet. This method used shared multi-layer perceptrons and max pooling for the feature extraction. Maron et al.~\cite{maron2017convolutional} applied CNN to sphere-type shapes using a global parametrization to a planar flat-torus.

Similar to considering images as an array of 2D pixels, discretizing 3D models into voxels is a good way to organize the shape information for CNN-based shape analysis. Wu et al.~\cite{wu20153d} proposed 3D ShapeNets for 3D object detection. They represented a 3D shape as a probability distribution of binary variables on voxels. Maturana and Scherer~\cite{maturana2015voxnet} used similar strategy to encode large point cloud datasets. They used a binary occupancy grid to distinguish free and occupied spaces, a density grid to estimate the probability that the voxel would block a sensor beam, and a hit grid to record the hit numbers.
Such volumetric discretization consumes memory cubically w.r.t. the voxel resolution, thus is not feasible for high-resolution shape analysis. Observing the fact that the spatial occupancy of 3D data is often sparse, Wang et al.~\cite{wang2015voting} designed a feature-centric voting algorithm named Vote3D for fast recognition of cars, pedestrians and bicyclists from the KITTI database~\cite{Geiger2013IJRR} using the sliding window method. More importantly, they demonstrated mathematical equivalence between the sparse convolution and voting. Based on this, Engelcke et al.~\cite{engelcke2017vote3deep} proposed a method called Vote3Deep converting the convolution into voting procedures, which can be simply applied to the non-empty voxels. However, with more convolution layers added to the CNN, this method quickly becomes prohibitive.

Octree-based data structures have been proven an effective way to reduce the memory consumption of 3D shapes. For example, Riegler et al.~\cite{riegler2017octnet} proposed a hybrid grid-octree data structure to support high-resolution 3D CNNs. Our work is most relevant to OCNN~\cite{OCNN_2017}, which used an octree to store the surface features of a 3D model and reduced the memory consumption for 3D CNNs to $\mathbf{O}(N^2)$. For the octree data structure, an octant is subdivided into eight children octants if it intersects with the model's surface regardless if all of those eight children octants are on the model. Therefore, an OCNN's subdivision also yields $\mathbf{O}(N^2)$ futile octants that do not contain useful features of the model. On the other hand, we use multi-level PSH~\cite{PSH_2006} to organize voxelized 3D models. PSH is nearly minimal while retaining an as cache-friendly as possible random access. As a result, the memory footprint of HCNN is close to the theoretic lower bound. Unlike in the original PSH work~\cite{PSH_2006}, the main hash table only stores the data index, and the real feature data is compactly assembled in a separate data array. We investigate how to seamlessly synergize hierarchical PSH-based models with CNN operations so that they can be efficiently executed on the GPU.

\section{Spatial Hashing for 3D CNN}
\label{sec:hash}
\begin{figure}[t!]
  \centering
  \includegraphics[width=\linewidth]{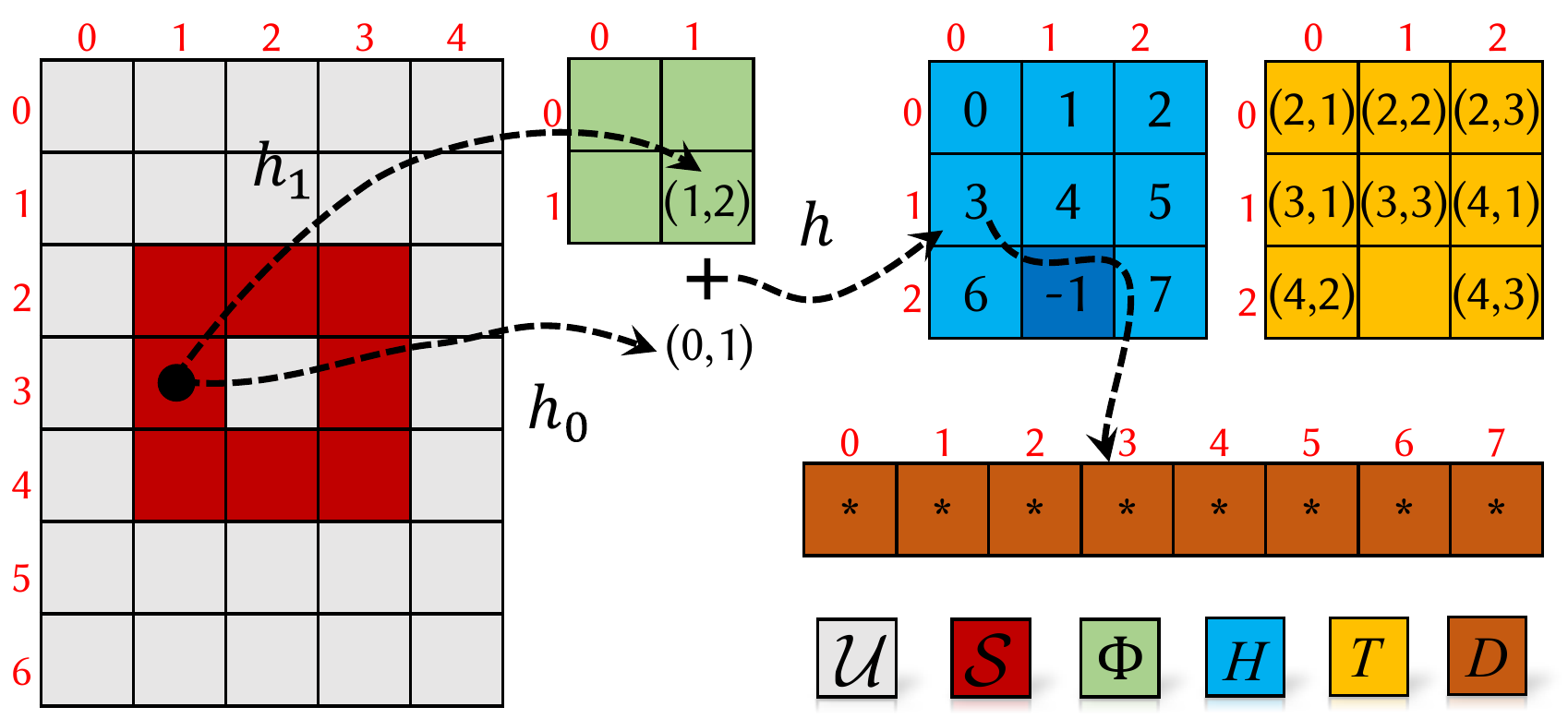}
  \caption{An illustrative 2D example of the constitution of our PSH. The domain $\mathscr{U}$ consists of $7\times5$ 2D voxels or pixels. The red-shaded pixels stand for the input model. The green, blue, yellow and brown tables are the offset table ($\Phi$), hash table ($H$), position tag ($T$) and data array ($D$) respectively.}\label{fig:psh}
\end{figure}

For a given input 3D shape, either a triangle soup/mesh or a point cloud, we first uniformly scale it to fit a unit sphere pivoted at the model's geometry center. Then, an axis-aligned bounding cube is built, whose dimension equals to the sphere's diameter.  Doing so ensures that the model remains inside the bounding box under arbitrary rotations, so that we can further apply the training data augmentation during the training (see e.g. \secref{subsec:classification}). This bounding cube is subdivided into grid cells or \emph{voxels} along $x$, $y$, and $z$ axes. A voxel is a small equilateral cuboid. It is considered non-empty when it encapsulates a small patch of the model's boundary surface. As suggested in~\cite{OCNN_2017}, we put extra sample points on this embedded surface patch, and the averaged normal of all the sample points is fed to the CNN as the input signal. For an empty voxel, its input is simply a zero vector.


\subsection{Multi-level PSH}
\label{subsec:multi_level_psh}
A set of hierarchical PSHs are built. At each level of the hierarchy, we construct a data array $D$, a hash table $H$, an offset table $\Phi$ and a position tag $T$. The data array at the finest level stores the input feature (i.e. normal direction of the voxel).
Let $\mathscr{U}$ be a $d$-dimensional discrete spatial domain with $u=\bar{u}^d$ voxels, out of which the sparse geometry data $\mathscr{S}$ occupies $n$ grid cells (i.e. $n=|\mathscr{S}|$). In other words, $\mathscr{U}$ represents all the voxels within the bounding cube at the given resolution, and $\mathscr{S}$ represents the set of voxels intersecting with the input model. We seek for a hash table $H$, which is a $d$-dimensional array of size $m=\bar{m}^d \geq n$ and a $d$-dimensional offset table $\Phi$ of size $r=\bar{r}^d$. By building maps $h_0 : p \rightarrow p \bmod \bar{m}$ from $\mathscr{U}$ to the hash table $H$ and $h_1 : p \rightarrow p \bmod \bar{r}$ from $\mathscr{U}$ on the offset table $\Phi$, one can obtain the perfect hash function mapping each non-empty voxel on the 3D shape $p \in \mathscr{S} $ to a unique slot $s = h(p)$ in the hash table as:
\begin{equation}\label{eq:hash_func}
h(p) =  h_0(p) + \Phi[h_1(p)] \bmod\bar{m}.
\end{equation}
Note that the hash table $H$ possesses slightly excessive slots (i.e. $m=\bar{m}^d \geq n$) to make sure that the hashing representation of $\mathscr{S}$ is collision free. A \texttt{NULL} value is stored at those redundant slots in $H$. Clearly, these \texttt{NULL} values should not participate in the CNN operations like batch normalization and scale. To this end, we assemble all the data for $\mathscr{S}$ into a compact $d$-dimensional array $D$ of size $n$. $H$ only houses the data index in $D$. If a slot in $H$ is redundant, it is indexed as $-1$ so that the associated data query is skipped.
\begin{figure*}[t!]
  \centering
  \includegraphics[width=0.9\linewidth]{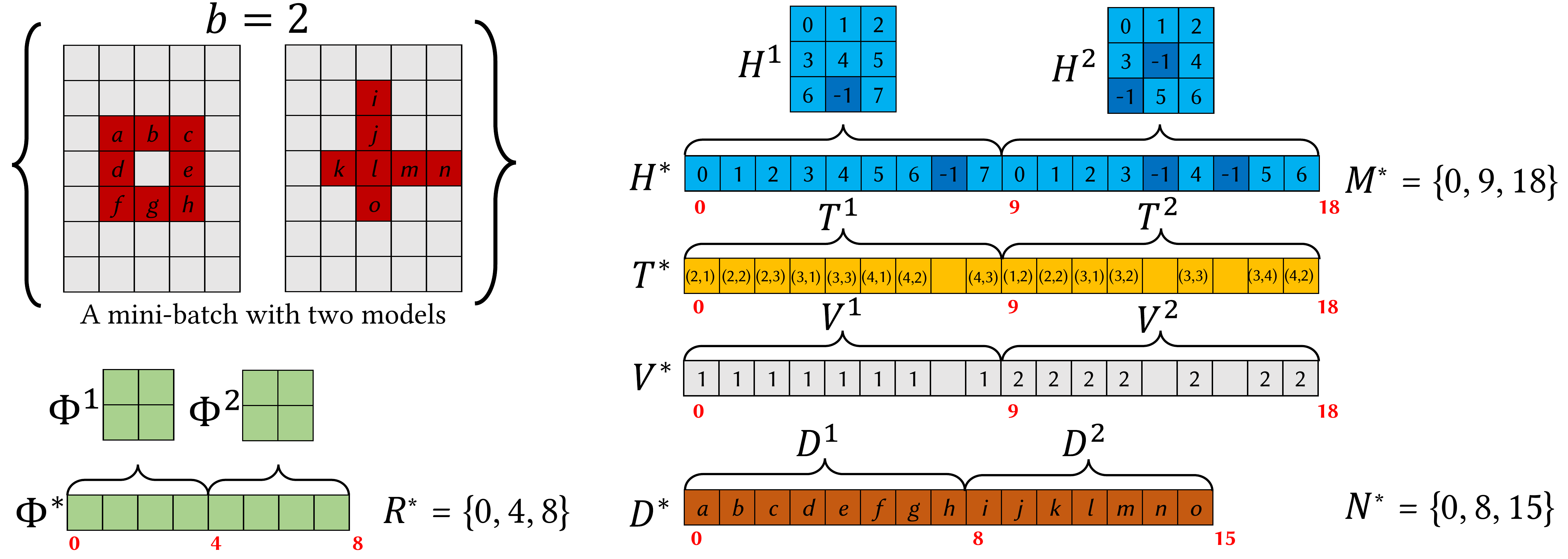}
  \caption{PSH data structures for a mini-batch of two models. All the feature data for the red-shaded pixels are stored in the super data array $D^*$, which consists of the data arrays of each individual models. Super hash table $H^*$, position tag $T^*$ and model index table $V^*$ are of the same size. For a give hash slot indexed at $i_{H^*}$, one can instantly know that this voxel, if not empty, is on the $V^*[i_{H^*}]$-th model in the batch by checking the model index table $V^*$. This information bridges the data sets from different hierarchy levels. With the auxiliary accumulated index tables $R^*$, $M^*$, and $N^*$, we can directly pinpoint the data using local index offset by $R^*[V^*[i_{H^*}]-1]$, $M^*[V^*[i_{H^*}]-1]$ and $N^*[V^*[i_{H^*}]-1]$ respectively. For instance in this simple example, when the local hash index is computed using \eqnref{eq:hash_func} for a non-empty voxel on the second model, its hash index in $H^*$ can then be obtained by offsetting the local hash index by $M^*[2-1]=9$.
  }\label{fig:minibatch}
\end{figure*}

Empty voxels (i.e. when $p\in\mathscr{U}\setminus\mathscr{S}$) may also be visited during CNN operations like convolution and pooling. Plugging these voxels' indices into \eqnref{eq:hash_func} is likely to return incorrect values that actually correspond to other non-empty grid cells. To avoid this mismatch, we adopt the strategy used in~\cite{PSH_2006} adding an additional position tag table $T$, which has the same size of $H$. $T[i]$ stores the voxel index for the corresponding slot at $H[i]$. Therefore when a grid cell $p$ is queried, we first check its data index in $H$ or $H[h(p)]$. If it returns a valid index other than -1, we further check the position tag $T[h(p)]$ to make sure $T[h(p)]=p$. Otherwise, $p\in\mathscr{U}\setminus\mathscr{S}$ is an off-model voxel and the associated CNN operation should be skipped. In our implementation, we use a 16-bit position tag for each $x$, $y$ and $z$ index, which supports the voxelization resolution up to $65,536^3$.

\figref{fig:psh} gives an illustrative 2D toy example. The domain $\mathscr{U}$ is a $7\times5$ 2D pixel grid. The red-shaded pixels stand for the input model, thus $n=|\mathscr{S}|=8$. We have a $3\times 3$ hash table $H$ (i.e. $\bar{m}=3$ and it is the blue table in the figure) and a $2\times 2$ offset table (i.e. $\bar{r}=2$ and it is the green table in the figure). Assume that the pixel $p(3,1)$ is queried and  $h_0$ yields $h_0(3,1)=(3 \bmod\bar{m},1\bmod\bar{m})=(0,1)$. $h_1(3,1)=(3 \bmod\bar{r},1\bmod\bar{r})=(1,1)$ gives the 2D index in the offset table. $\Phi(1,1)=(1,2)$, which is added to $h_0(p)$ to compute the final index in H: $\Phi(1,1)+h_0(p)=(1+0 \bmod\bar{m}, 2+1\bmod\bar{m})=(1,0)$. Before we access the corresponding data cell in $D$ (the fourth cell in this example because $H(1,0)=3$), the position tag table (the yellow table) is queried. Since $T(1,0)=(3,1)$, which equals to the original pixel index of $p$, we know that $p\in\mathscr{S}$ is indeed on the input model. Note that in this example, $H(2,1)$ is a redundant slot (colored in dark blue in \figref{fig:psh}). Therefore, the corresponding index is -1.

\subsection{Mini-batch with PSH}
\label{subsec:minibatch}
During the CNN training, it is typical that the network parameters are optimized over a subset of the training data, referred to as a mini-batch. Let $b$ be the batch size and $l$ be the resolution level. In order to facilitate per-batch CNN training, we build a ``super-PSH'' by attaching $H$, $\Phi$, $T$ for all the models in a batch: $H_l^* = \big\{H_l^1,H_l^2,...,H_l^b\big\}$, $\Phi_l^* = \big\{\Phi_l^1,\Phi_l^2,...,\Phi_l^b\big\}$, and $T_l^* = \big\{T_l^1,T_l^2,...,T_l^b\big\}$ as illustrated in \figref{fig:minibatch}. That is we expand each of these $d$-dimensional tables into a 1D array and concatenate them together. The data array $D^*_l$ of the batch is shaped as a row-major $c_l$ by $\sum_{i=1}^b|\mathscr{S}_l^i|$ matrix, where $c_l$ is the number of channels at level $l$, and  $\sum_{i=1}^b|\mathscr{S}_l^i|$ is the total number of non-empty voxels of all the models in the batch. A column of $D^*_l$ is a $c_l$-vector, and it stores the features of the corresponding voxel. The dimensionality of $H^i_l$, $\Phi^i_l$, and $D^i_l$ is also packed as $\bar{m}_l^*=\big\{\bar{m}_l^1,\bar{m}_l^2,...,\bar{m}_l^b\big\}$, $\bar{r}_l^*=\big\{\bar{r}_l^1,\bar{r}_l^2,...,\bar{r}_l^b\big\}$, and  $n_l^*=\big\{n_l^1,n_l^2,...,n_l^b\big\}$.

%

In addition, we also record accumulated indices for $H$, $\Phi$ and $D$ as: $M_l^* = \big\{0,M_l^1,M_l^2,...,M_l^b\big\}$, $R_l^* = \big\{0,R_l^1,R_l^2,...,R_l^b\big\}$ and $N_l^* = \big\{0,N_l^1,N_l^2,$ $...,N_l^b\big\}$ where
\begin{equation*}
M_l^i=\sum_{k=1}^{i}(\bar{m}_l^k)^d=\sum_{k=1}^{i}{m_l^k},\quad
R_l^i=\sum_{k=1}^{i}(\bar{r}_l^k)^d=\sum_{k=1}^{i}{r_l^k},\quad
N_l^i=\sum_{k=1}^{i}{n_l^k}.
\end{equation*}
Indeed, $M_l^*$, $R_l^*$ and $N_l^*$ store the super table (i.e. $H^*_l$, $\Phi^*_l$, $T^*_l$, and $D^*_l$) offsets of the $k$-th model in the batch. For instance, the segment of $H^*_l$ starting from $H^*_l\big[M^*_l[k-1]\big]$ to $H^*_l\big[M^*_l[k]-1\big]$ corresponds to the hash table $H^k_l$; the segment from $\Phi^*_l\big[R^*_l[k-1]\big]$
to $\Phi^*_l\big[R^*_l[k]-1\big]$ corresponds to the offset table $\Phi^k_l$; the segment from $T^*_l\big[M^*_l[k-1]\big]$
to $T^*_l\big[M^*_l[k]-1\big]$ corresponds to the position tag $T^k_l$; and the segment from $D^*_l\big[N^*_l[k-1]\big]$
to $D^*_l\big[N^*_l[k]-1\big]$ is the data array $D^k_l$. Lastly, we build a model index table $V_l^* = \big\{V_l^1,V_l^2,...,V_l^b\big\}$  for the inverse query. Here, $V_l^i$ has the same size as $H_l^i$ does, and each of its slots stores the model's index in a batch: $V_l^i(\cdot) = i$.

\section{CNN Operations with Multi-level PSH}
\label{sec:CNN_operations}
In this section we show that how to apply CNN operations like convolution/transposed convolution, pooling/unpooling, batch normalization and scale to the PSH-based data structure so that they can be efficiently executed on the GPU.
\vspace{5 pt}
\noindent\textbf{Convolution}\hspace{5 pt}
The convolution operator $\Psi_c$ in the unrolled form is:
\begin{equation}
\Psi_c(p) = \sum_{n}\sum_{i}\sum_{j}\sum_{k}{W_{ijk}^{(n)}\cdot \mathcal{T}^{(n)}(p_{ijk})},
\label{eq:conv}
\end{equation}
where $p_{ijk}$ is a neighboring voxel of voxel $p\in\mathscr{S}$. $\mathcal{T}^{(n)}$ and $W_{ijk}^{(n)}$ are the feature vector and the kernel weight of the $n$-th channel. This nested summation can be reshaped as a matrix product~\cite{chellapilla2006high} and computed efficiently on the GPU:
\begin{equation}
\mathbf{D}_o = \mathbf{W}\cdot\widetilde{\mathbf{D}}_i.
\label{eq:mat_conv}
\end{equation}
Let $l_i$ and $l_o$ denote the input and output hierarchy levels of the convolution. $\mathbf{D}_o$ is essentially the matrix representation of the output data array $D^*_{l_o}$. Each column of $\mathbf{D}_o$ is the feature signal of an output voxel. A row vector in $\mathbf{W}$ concatenates vectorized kernel weights for all the input channels, and the number of rows in $\mathbf{W}$ equals to the number of convolution kernels employed. We design a subroutine $\code{hash2col}$ to assist the assembly of matrix $\widetilde{\mathbf{D}}_i$, which fetches feature values out of the input data array $D^*_{l_i}$ so that a column of $\widetilde{\mathbf{D}}_i$ stacks feature signals within the receptive fields covered by kernels.

\begin{algorithm}[t!]
\footnotesize
\SetNlSty{tiny}{}{:}
\KwIn{$b$, $c_{l_i}$, $D^*_{l_i}$, $H^*_{l_i}$, $M^*_{l_i}$, $R^*_{l_i}$, $N^*_{l_i}$ $V^*_{l_o}$, $T^*_{l_o}$, $H^*_{l_o}$, $M^*_{l_o}$, $N^*_{l_o}$, $F$, $S_s$, $S_p$}
\KwOut{$\widetilde{\mathbf{D}}_i$}
\nonl\hrulefill\\
launch $c_{l_i}\cdot M^*_{l_o}[b]$ threads; \\
\tcc{$i_{thrd}$ is the thread index}
\For {$i_{thrd}=0:c_{l_i}\cdot M^*_{l_o}[b]-1$}
{
    $i_c\leftarrow \big\lfloor i_{thrd}/M^*_{l_o}[b]\big\rfloor$ \tcp*{$i_c$ is the channel index}
    $i_{H^*_{l_o}}\leftarrow i_{thrd}-i_c\cdot M^*_{l_o}[b]$;\\
    $v\leftarrow V^*_{l_o}[i_{H^*_{l_o}}]$\tcp*{$v$ is the model index in the mini-batch}
    $col\leftarrow N^*_{l_o}\big[v-1\big]+H_{l_o}^*[i_{H^*_{l_o}}]$\tcp*{$col$ is the column index}
    \If {$H^*_{l_o}[i_{H^*_{l_o}}] = -1$}
    {
        return \tcp*{$i_{H^*_{l_o}}$ points to an empty hash slot}
    }
    \Else
    {
        $p_{l_o}\leftarrow T^*_{l_o}[i_{H^*_{l_o}}]$\tcp*{$p_{l_o}$ is the voxel position}
        $\mathscr{R}_{l_i}\leftarrow\emptyset$\tcp*{$\mathscr{R}_{l_i}$ is the receptive field on $\mathscr{U}_{l_i}$}
        \tcc{$S_s$ and $S_p$ are the stride size and padding size}
        \If {$S_s = 1$}
        {
            $\mathscr{R}_{l_i}\leftarrow\mathscr{U}_{l_i}[p_{l_o}-(F-1)/2, p_{l_o}+(F-1)/2]^3$;\\
        }
        \Else
        {
            $\mathscr{R}_{l_i}\leftarrow\mathscr{U}_{l_i}[p_{l_o}\cdot S_s - S_p, p_{l_o}\cdot S_s - S_p + F-1]^3$;\\
        }
        $row \leftarrow 0$\tcp*{$row$ is current row index in $\widetilde{\mathbf{D}}_i$}
        \tcc{iterate all the voxels on the receptive field}
        \For {$p_{l_i}\in\mathscr{R}_{l_i}$}
        {
            $i_{\Phi^*_{l_i}}\leftarrow R^*_{l_i}[v-1]+h_1(p_{l_i})$;\\
            $i_{H^*_{l_i}}\leftarrow M^*_{l_i}[v-1] + \big(h_0(p_{l_i}) + \Phi^*_{l_i}[i_{\Phi^*_{l_i}}]\bmod \bar{m}_{l_i}$\big);\\
            \If {$H^*_{l_i}[i_{H^*_{l_i}}]\neq-1$ and $p_{l_i}=T^*_{l_i}[i_{H^*_{l_i}}]$}
            {
                $i_{D^*_{l_i}}\leftarrow N^*_{l_i}[v-1] + H^*_{l_i}[i_{H^*_{l_i}}]$;\\
                $\widetilde{\mathbf{D}}_i[i_c\cdot F + row, col]\leftarrow D^*_{l_i}[i_c, i_{D^*_{l_i}}]$;\\
            }
            \Else
            {
                $\widetilde{\mathbf{D}}_i[i_c\cdot F + row, col]\leftarrow 0$;\\
            }
            \tcc{assume $p_{l_i}$ is iterated according to its spatial arrangement in $\mathscr{R}_{l_i}$}
            $row\leftarrow row + 1$;\\
         }
    }
}
\caption{\code{hash2col} subroutine}
\label{alg:hash2col}
\end{algorithm}

The algorithmic procedure for $\code{hash2col}$ is detailed in \lstref{alg:hash2col}. In practice, we launch $c_{l_i}\cdot M^*_{l_o}[b]$ \texttt{CUDA} threads in total, where $c_{l_i}$ is the number of input channels. Recall that $M^*_{l_o}[b]$ is the last entry of the accumulated index array $M_{l_o}^*$ such that $M^*_{l_o}[b]=M_{l_o}^b$, and it gives the total number of hash slots on $H^*_{l_o}$.
Hence, our parallelization scheme can be understood as assigning a thread to collect necessary feature values within the receptive field for each output hash slot per channel. The basic idea is to find the receptive field $\mathscr{R}_{l_i}\subset\mathscr{U}_{l_i}$ that corresponds to an output voxel $p_{l_o}$ and retrieve features for $\widetilde{\mathbf{D}}_i$. A practical challenge lies in the fact that output and input data arrays may reside on the voxel grids of different hierarchy levels. Therefore, we need to resort to the PSH mapping (\eqnref{eq:hash_func}) and the position tag table to build the necessary output-input correspondence.

Given a thread index $i_{thrd}$ ($0\leq i_{thrd} \leq c_{l_i}\cdot M^*_{l_o}[b]-1$), we compute its associated channel index $i_c$ as $i_c=\big\lfloor i_{thrd}/M^*_{l_o}[b]\big\rfloor$. Its super hash index $i_{H^*_{l_o}}$ (i.e. the index in $H^*_{l_o}$) is simply $i_{H^*_{l_o}} = i_{thrd} - i_c\cdot M^*_{l_o}[b]$, so that we know that this thread is for the $V^*_{l_o}[i_{H^*_{l_o}}]$-th model in the batch (recall that $V^*_{l_o}$ is the model index table). If $H^*_{l_o}[i_{H^*_{l_o}}]\neq-1$ meaning this thread corresponds to a valid non-empty voxel, the index of the column in $\mathbf{D}_o$ that houses the corresponding output feature is $N^*_{l_o}\big[V^*_{l_o}[i_{H^*_{l_o}}]-1\big]+H_{l_o}^*[i_{H^*_{l_o}}]$.

With the help of the position tag table $T_{l_o}^*$, the index of the output voxel in $\mathscr{U}_{l_o}$ associated with the thread $i_{thrd}$ can be retrieved by $p_{l_o} = T_{l_o}^*[i_{H^*_{l_o}}]$, based on which we can obtain the input voxel positions within the receptive field and construct the corresponding column in $\widetilde{\mathbf{D}}_i$. Specifically, if the stride size is one, indicating the voxel resolution is unchanged after the convolution or $l_i=l_o$,
the input model has the same hash structure as the output. In this case, the receptive field associated with $p_{l_o}$ spans from $p_{l_o}-(F-1)/2$ to $p_{l_o}+(F-1)/2$ along each dimension on $\mathscr{U}_{l_i}$ denoted as $\mathscr{U}_{l_i}[p_{l_o}-(F-1)/2, p_{l_o}+(F-1)/2]^d$. Here, $F$ is the kernel size. On the other hand, if the stride size is larger than one, the convolution will down-sample the input feature, and the receptive field on $\mathscr{U}_{l_i}$ is $\mathscr{U}_{l_i}[p_{l_o}\cdot S_s - S_p, p_{l_o}\cdot S_s - S_p+F-1]^d$ with the stride size $S_s$ and the padding size $S_p$. For irregular kernels~\cite{dai2017deformable,ma2017irregular}, we can similarly obtain the corresponding receptive field on $\mathscr{U}_{l_i}$ based on $p_{l_o}$.




As mentioned, for a non-empty voxel $p_{l_i}\in\mathscr{U}_{l_i}$ within the receptive field of a given output voxel $p_{l_o}\in\mathscr{U}_{l_o}$, we know that it belongs to the $v$-th model of the batch, where $v=V^*_{l_o}[i_{H^*_{l_o}}]$. Therefore, its offset index in $\Phi^*_{l_i}$ can be computed as:
\begin{equation}\label{eq:super_offset_idx}
i_{\Phi^*_{l_i}}=R^*_{l_i}[v-1]+h_1(p_{l_i}),
\end{equation}
where $R^*_{l_i}$ is the accumulated offset index array at level $l_i$, and $R^*_{l_i}[v-1]$ returns the starting index of the offset table $\Phi^v_{l_i}$ in the super table $\Phi^*_{l_i}$. $h_1(p_{l_i})$ computes the (local) offset index. Thus, the offset value of $p_{l_i}$ can be queried by $\Phi^*_{l_i}[i_{\Phi^*_{l_i}}]$. The index of $p_{l_i}$ in the super hash table $H^*_{l_i}$ can be computed similarly as:
\begin{equation}\label{eq:super_hash_idx}
{\setlength\arraycolsep{0 pt}
\begin{array}{rl}
i_{H^*_{l_i}}&=M^*_{l_i}[v-1] + \big(h_0(p_{l_i}) + \Phi^*_{l_i}[i_{\Phi^*_{l_i}}]\bmod \bar{m}_{l_i}\big)\\
&=M^*_{l_i}[v-1] + \big( h_0(p_{l_i}) + \Phi^*_{l_i}[R^*_{l_i}[v-1]+h_1(p_{l_i})]\bmod \bar{m}_{l_i}\big).
\end{array}}
\end{equation}
Here, $h_0(p_{l_i})$ and $h_1(p_{l_i})$ are maps defined on hierarchy level $l_i$. If $H^*_{l_i}[i_{H^*_{l_i}}]\neq-1$ and the position tag is also consistent (i.e. $T^*_{l_i}[i_{H^*_{l_i}}]=p_{l_i}$), we fetch the feature from the data array by $D^*_{l_i}[i_{D^*_{l_i}}]$, where
\begin{equation}\label{eq:data_idx}
i_{D^*_{l_i}}=N^*_{l_i}[v-1] + H^*_{l_i}[i_{H^*_{l_i}}].
\end{equation}
Otherwise, a zero value is returned.

%


\begin{algorithm}[t!]
\footnotesize
\SetNlSty{tiny}{}{:}
\KwIn{$b$, $c_{l_i}$, $\delta\widetilde{\mathbf{D}}_i$, $H^*_{l_i}$, $M^*_{l_i}$, $R^*_{l_i}$, $N^*_{l_i}$ $V^*_{l_o}$, $T^*_{l_o}$, $H^*_{l_o}$, $M^*_{l_o}$, $N^*_{l_o}$, $F$, $S_s$, $S_p$}
\KwOut{$\delta D^*_{l_i}$}
\nonl\hrulefill\\
$\delta D^*_{l_i}[:]\leftarrow 0$\tcp*{all the entries in $\delta D^*_{l_i}$ are initialized as 0}
launch $c_{l_i}\cdot M^*_{l_o}[b]$ threads; \\
\tcc{$i_{thrd}$ is the thread index}
\For {$i_{thrd}=0:c_{l_i}\cdot M^*_{l_o}[b]-1$}
{
    $i_c\leftarrow \big\lfloor i_{thrd}/M^*_{l_o}[b]\big\rfloor$ \tcp*{$i_c$ is the channel index}
    $i_{H^*_{l_o}}\leftarrow i_{thrd}-i_c\cdot M^*_{l_o}[b]$;\\
    $v\leftarrow V^*_{l_o}[i_{H^*_{l_o}}]$\tcp*{$v$ is the model index in the mini-batch}
    $col\leftarrow N^*_{l_o}\big[v-1\big]+H_{l_o}^*[i_{H^*_{l_o}}]$\tcp*{$col$ is the column index}
    \If {$H^*_{l_o}[i_{H^*_{l_o}}] = -1$}
    {
        return \tcp*{$i_{H^*_{l_o}}$ points to an empty hash slot}
    }
    \Else
    {
        $p_{l_o}\leftarrow T^*_{l_o}[i_{H^*_{l_o}}]$\tcp*{$p_{l_o}$ is the voxel position}
        $\mathscr{R}_{l_i}\leftarrow\emptyset$\tcp*{$\mathscr{R}_{l_i}$ is the receptive field on $\mathscr{U}_{l_i}$}
        \tcc{$S_s$ and $S_p$ are the stride size and padding size}
        \If {$S_s = 1$}
        {
            $\mathscr{R}_{l_i}\leftarrow\mathscr{U}_{l_i}[p_{l_o}-(F-1)/2, p_{l_o}+(F-1)/2]^3$;\\
        }
        \Else
        {
            $\mathscr{R}_{l_i}\leftarrow\mathscr{U}_{l_i}[p_{l_o}\cdot S_s - S_p, p_{l_o}\cdot S_s - S_p + F-1]^3$;\\
        }
        $row \leftarrow 0$\tcp*{$row$ is current row index in $\delta\widetilde{\mathbf{D}}_i$}
        \tcc{iterate all the voxels on the receptive field}
        \For {$p_{l_i}\in\mathscr{R}_{l_i}$}
        {
            $i_{\Phi^*_{l_i}}\leftarrow R^*_{l_i}[v-1]+h_1(p_{l_i})$;\\
            $i_{H^*_{l_i}}\leftarrow M^*_{l_i}[v-1] + h_0(p_{l_i}) + \big( \Phi^*_{l_i}[i_{\Phi^*_{l_i}}]\bmod \bar{m}_{l_i}\big)$;\\
            \If {$H^*_{l_i}[i_{H^*_{l_i}}]\neq-1$ and $p_{l_i}=T^*_{l_i}[i_{H^*_{l_i}}]$}
            {
                $i_{D^*_{l_i}}\leftarrow N^*_{l_i}[v-1] + H^*_{l_i}[i_{H^*_{l_i}}]$;\\
                $\delta D^*_{l_i}[i_c, i_{D^*_{l_i}}]\leftarrow \delta D^*_{l_i}[i_c, i_{D^*_{l_i}}]+\delta\widetilde{\mathbf{D}}_i[i_c\cdot F + row, col]$;
            }
            $row\leftarrow row + 1$;\\
         }
    }
}
\caption{\code{col2hash} subroutine}
\label{alg:col2hash}
\end{algorithm}

\vspace{5 pt}
\noindent\textbf{Back propagation \& weight update}\hspace{5 pt}
During the CNN training and optimization, the numerical gradient of kernels' weights is computed as:
\begin{equation}\label{eq:weight_gradient}
\delta\mathbf{W}=\delta\mathbf{D}_o\cdot\widetilde{\mathbf{D}}_i^\top,
\end{equation}
where $\delta\mathbf{D}_o$ is the variation of the output data array $\mathbf{D}_o$. In order to apply \eqnref{eq:weight_gradient} in previous CNN layers, we also calculate how the variational error is propagated back:
\begin{equation}\label{eq:back_propagation}
\delta\widetilde{\mathbf{D}}_i=\mathbf{W}^\top\cdot\delta\mathbf{D}_o.
\end{equation}
Clearly, we need to re-pack the errors in $\delta\widetilde{\mathbf{D}}_i$ in accordance with the format of the data array $D^*_{l_i}$ so that the resulting matrix $\delta\mathbf{D}_i$ can be sent to the previous CNN layer. This process is handled by the \code{col2hash} subroutine, outlined in \lstref{alg:col2hash}. As the name implies, \code{col2hash} is quite similar to \code{hash2col} except at line 26, where variational errors from the receptive field is lumped into a single accumulated error.



\vspace{5 pt}
\noindent\textbf{Pooling, unpooling \& transposed convolution}\hspace{5 pt}
The pooling layer condenses the spatial size of the input features by using a single representative activation for a receptive field. This operation can be regarded as a special type of convolution with a stride size $S_s>1$. Therefore, \code{hash2col} subroutine can also assist the pooling operation. The average-pooling is dealt with as applying a convolution kernel with all the weights equal to $1/F^3$. For the max-pooling, instead of performing a stretched inner product across the receptive field, we output the maximum signal after the traversal of the receptive field (the \texttt{for} loop at line 20 in \lstref{alg:hash2col}). Unlike OCNN~\cite{OCNN_2017}, our framework supports any stride sizes for the pooling since the PSH can be generated on the grid of an arbitrary resolution.

The unpooling operation aims to partially revert the input activation after the pooling, which could be useful for understanding the CNN features~\cite{zeiler2011adaptive,zeiler2014visualizing} or restoring the spatial structure of the input activations for segmentation~\cite{noh2015learning}, flow estimation~\cite{dosovitskiy2015flownet}, and generative modeling~\cite{noh2015learning}. During the max-pooling, we record the index of the maximum activation for each receptive field (known as the \emph{switch}). When performing the max-unpooling, the entire receptive field corresponding to an input voxel is initialized to be zero, and the feature signal is restored only at the recorded voxel index. The average-unpooling is similarly handled, where we evenly distribute the input activation over its receptive field.

Transposed convolution is also referred to as deconvolution or fractionally strided convolution~\cite{zhu2010learning}, which has been proven useful for enhancing the activation map~\cite{zeiler2011adaptive,shi2016real}. Mathematically, the transposed convolution is equivalent to the regular convolution and can be dealt with using \code{hash2col} subroutine. However, doing so involves excessive zero padding and thus degenerates network's performance. In fact, the deconvolution flips the input and output of the forward convolution using a transposed kernel as: $\widetilde{\mathbf{D}}_o=\mathbf{W}^\top\cdot\mathbf{D}_i$, which is exactly how we handle the error back propagation (i.e. \eqnref{eq:back_propagation}). Therefore, the \code{col2hash} subroutine can be directly used for deconvolution operations.

\vspace{5 pt}
\noindent\textbf{Other CNN operations}\hspace{5 pt}
Because all the feature values in HCNN are compactly stored in the data array $D^*$, operations that are directly applied to the feature values like batch normalization~\cite{ioffe2015batch} and scale can be trivially parallelized on GPU.

\section{Experimental Results}
\label{sec:experiment}

Our framework was implemented on a desktop computer equipped with an \texttt{Intel I7-6950X} CPU ($3.0$ GHz) and an \texttt{nVidia GeForce 1080 Pascal} GPU with 8 GB DDR5 memory. We used \texttt{Caffe} framework~\cite{jia2014caffe} for the CNN implementation. The 3D models used are from \texttt{ModeNet40}~\cite{wu20153d} and \texttt{ShapeNet Core55}~\cite{chang2015shapenet}. Both are publicly available. The source code of HCNN can be found in the accompanying supplementary file. The executable and some of the training data in PSH format ($4.4$ GB in total) can also be downloaded via the anonymous \texttt{Google} \texttt{Drive} link, which can also be found in the supplementary file. We encourage readers to test HCNN by themselves.

\vspace{5 pt}
\noindent\textbf{Model rectification}\hspace{5 pt}
It has been noticed that normal informations on 3D models from the \texttt{ModeNet} database are often incorrect or missing. We fix the normal information by casting rays from 14 virtual cameras (at six faces and eight corners of the bounding cube). Some 3D models use a degenerated 2D plane to represent a thin shape. For instance, the back of a chair model may only consist of two flat triangles. To restore the volumetric information of such thin geometries, we displace the sample points on the model towards its normal direction by $1/(2\cdot\bar{u}_{max})$, where $\bar{u}_{max}$ denotes the voxel resolution at the finest hierarchy level. In other words, the model's surface is slightly dilated by a half-voxel size.

\subsection{Network Architecture}
\label{subsec:network}
A carefully fine-tuned network architecture could significantly improve the CNN result and relieve the training efforts. Nevertheless, this is neither the primary motivation nor the contribution of this work. In order to report an apple-to-apple comparison with peers and benchmark our method objectively, we employ a network similar to the well-known LeNet~\cite{lecun1998gradient}.

\begin{figure}[t!]
  \centering
  \includegraphics[width=\linewidth]{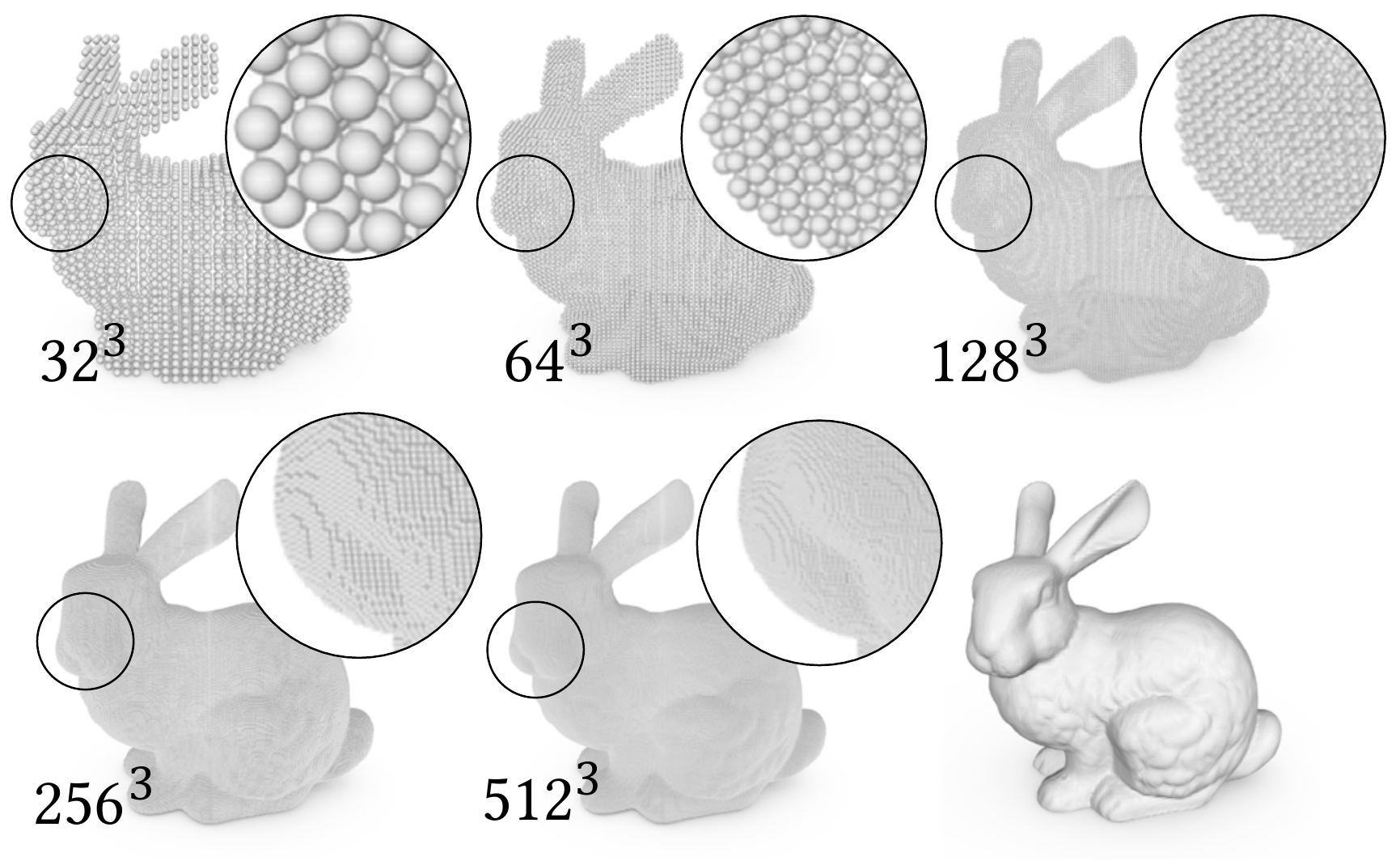}
  \caption{The benefit of dense voxelization is obvious. The discretized model better captures the geometry of the original shape at higher resolutions.
  }\label{fig:bunny}
\end{figure}

In our framework, the convolution and pooling operations are repeated from the finest level, and ReLU is used as the activation function. A batch normalization (BN) is also applied to reduce the internal covariance shift~\cite{ioffe2015batch}. Our PSH hierarchy allows very dense voxelization at the resolution of $512^3$ (i.e. see \figref{fig:bunny}), where the hierarchy level $l=9$. Each coarser level reduces the resolution by half, and the coarsest level has the resolution of $4^3$, where $l=2$. Such multi-level PSH configuration exactly matches the OCNN hierarchy, which allows us to better evaluate the performance between these two data structures. At each level, we has the same operation sequence as: $Convolution\rightarrow BN\rightarrow ReLU \rightarrow Pooling$. The receptive field of kernels is $3\times3\times3$, and the number of channels at the $l$-th level is set as $\max\{2,2^{9-l}\}$.

Three classic shape analysis tasks namely shape classification, retrieval, and segmentation are benchmarked. For the classification, two fully connected (FC) layers, a softmax layer and two dropout layers~\cite{hinton2012improving,srivastava2014dropout} ordered as: $Dropout\rightarrow FC(128)\rightarrow Dropout \rightarrow FC(N_c)\rightarrow Softmax \rightarrow Output$ are appended. Here, $FC(K)$ indicates $K$ neurons are set at the FC layer. For the shape retrieval, we use the output from the object classification as the key to search for the most similar shapes to the query. For the segmentation, we follow the DeconvNet~\cite{noh2015learning} structure, which adds a deconvolution network after a convolution network for dense predictions. The deconvolution network simply reverses the convolution procedure where the convolution and pooling operators are replaced by the deconvolution and unpooling operators. Specifically, at each level we apply $Unpooling\rightarrow Deconvolution \rightarrow BN \rightarrow ReLU$ and then move to the next finer level.

The reader may notice that our experiment setting transplants the one used in~\cite{OCNN_2017} except that all the features are organized using PSH rather than octrees. This is because we consider OCNN~\cite{OCNN_2017} as our primary competitor and would like the report an objective side-by-side comparison with it. Lastly, we would like to remind the reader again that HCNN is not restricted to power-of-two resolution changes. To the best of our knowledge, our HCNN is compatible with all the existing CNN architectures and operations.

\vspace{5 pt}
\noindent\textbf{Training specifications}\hspace{5 pt}
The network is optimized using the stochastic gradient
descent method. We set momentum as $0.9$ and weight decay as $0.0005$. A mini-batch consists of $32$ models. The dropout ratio is $0.5$. The initial learning rate is $0.1$, which is attenuated by a factor of 10 after 10 epochs.

\subsection{PSH Construction}
\label{subsec:psh_preconstruction}
As the data pre-processing, we construct a multi-level PSH for each 3D model. 
The size of the hash table is set as the smallest value satisfying $\bar{m}^3>|\mathscr{S}|$. Each hash table slot is an \texttt{int} type, which stores the data array index of $D$. Therefore, the hash table supports the high-resolution models up to $|\mathscr{S}|=2^{31}$, which is sufficient in our experiments. Next, we seek to make the offset table as compact as possible. The table size $\bar{r}$ is initialized as the smallest integer such that $\bar{r}^3\geq\sigma|\mathscr{S}|$ with the factor $\sigma$ empirically set as $\sigma=1/{2d}$, as in~\cite{PSH_2006}. An offset table cell is of 24 bits ($d\times8$), and each offset value is a $8$-bit \texttt{unsigned} \texttt{char}, which allows an offset up to $255$ at each dimension. If the hash construction fails, we increase $\bar{r}$ by $\sqrt[3]{2}$ (i.e. double the offset table capacity) until construction succeeds.
We refer readers to~\cite{PSH_2006} for implementation details. The construction of PSH is a pre-process and completely offline, yet it could be further accelerated on GPUs as~\cite{alcantara2009real}.



\subsection{Memory Analysis}
\label{subsec:memory_analysis}
An important advantage of using PSH is its excelling memory performance over state-of-the-art methods.
Our closest competitor is OCNN~\cite{OCNN_2017}, where the total number of the octants at the finest level does not depend on whether leaf octants intersect with the input model. Instead, it is determined by the occupancy of its parent: when the parent octant overlaps with the model's boundary, all of its eight children octants will be generated. While OCNN's memory consumption is quadratically proportional to the voxel resolution in the asymptotic sense, it also wastes $\mathbf{O}(N^2)$ memory for leaf octants that are not on the model. On the other hand, the memory overhead of our PSH-based data structure primarily comes from the difference between the actual model size i.e. the number of voxels on the model at the finest level and the hash table size (the offset tables are typically much smaller than the main hash table). Assume that the input model size is $|\mathscr{S}|=N^2$. The hash table size is $\bar{m}=\lceil N^{\frac{2}{3}}\rceil$, which is the smallest integer satisfying $\bar{m}^3>N^2$. By splitting $\lceil N^{\frac{2}{3}}\rceil$ as:
$\lceil N^{\frac{2}{3}}\rceil=N^{\frac{2}{3}}+\Delta M$, $0\leq\Delta M\leq1$, the memory overhead of PSH can then be estimated via:
\begin{equation}\label{eq:memory2}
\lceil N^{\frac{2}{3}}\rceil^3-N^2=\big(N^{\frac{2}{3}}+\Delta M\big)^3-N^2 \propto \Delta M N^{\frac{4}{3}},
\end{equation}
which is $\mathbf{O}(N^{\frac{4}{3}})$. In other words, the memory overhead of our HCNN is \emph{polynomially smaller} than OCNN.
\begin{figure}[h!]
  \centering
  \includegraphics[width=\linewidth]{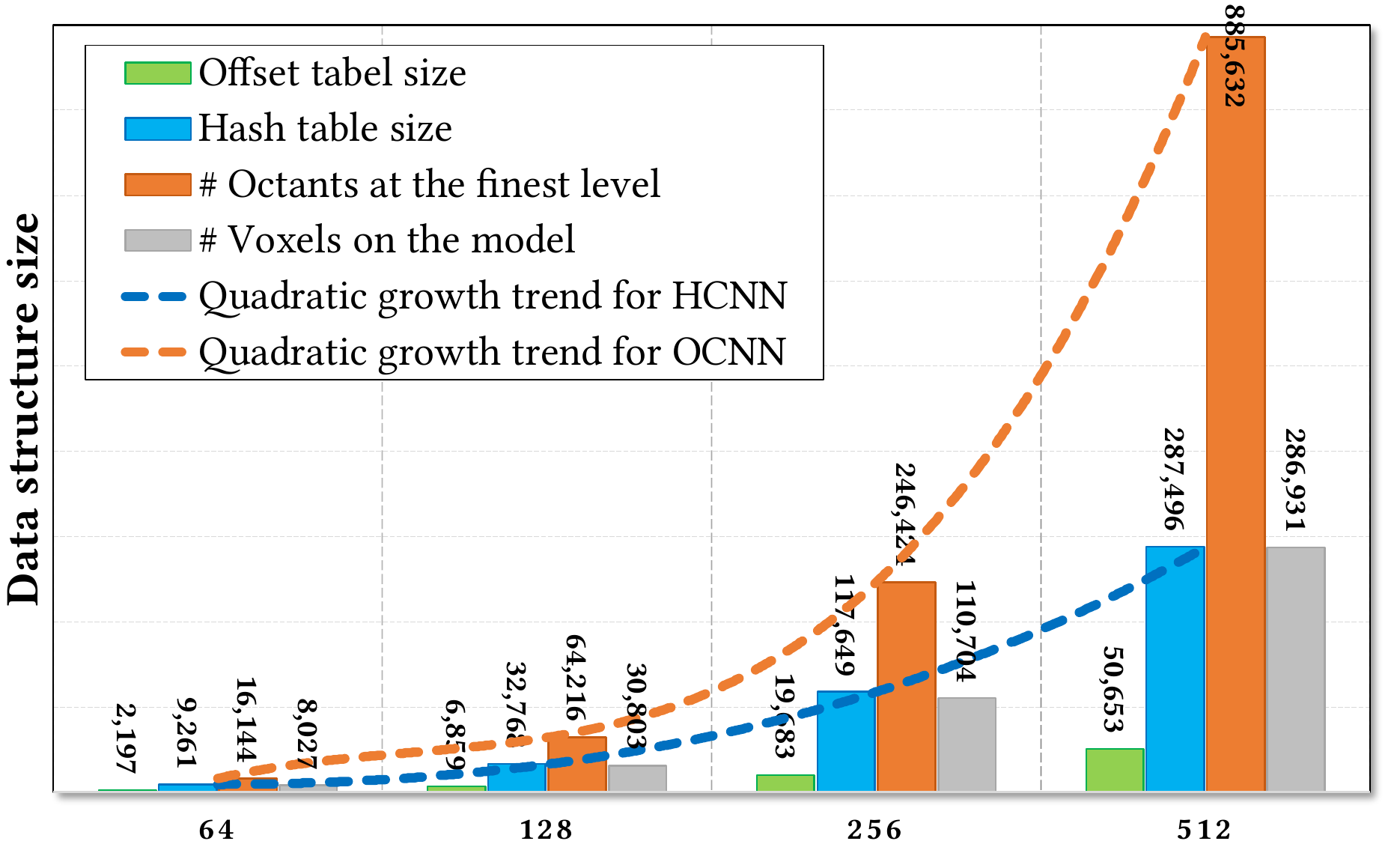}
  \caption{The sizes of PSH and octree data structures used to encode the bunny model under resolutions of $64^3$, $128^3$, $256^3$ and $512^3$. Under each resolution, the total number of octants and the sizes of the hash table ($H$) and offset table ($\Phi$) are reported. Their quadratic growth trends are also plotted.}\label{fig:memory_bunny}
\end{figure}

\figref{fig:memory_bunny} compares the sizes of the primary data structure for the bunny model (\figref{fig:bunny}) using OCNN and HCNN -- the total number of leaf octants and the size of the hash table ($H$) at the finest level. The size of the offset table $\Phi$ is typically an order smaller than $H$. Besides, the number of voxels on the model is also reported. It can be clearly seen from the figure that the size of the hash table is very close to the actual model size (i.e. the lower bound of the data structure). The latter is highlighted as grey bars in the figure. The asymptotic spatial complexity of both HCNN and OCNN are $\mathbf{O}(N^2)$, however the plotted growth trends show that HCNN is much more memory efficient than OCNN.

\begin{figure}[h!]
  \centering
  \includegraphics[width=\linewidth]{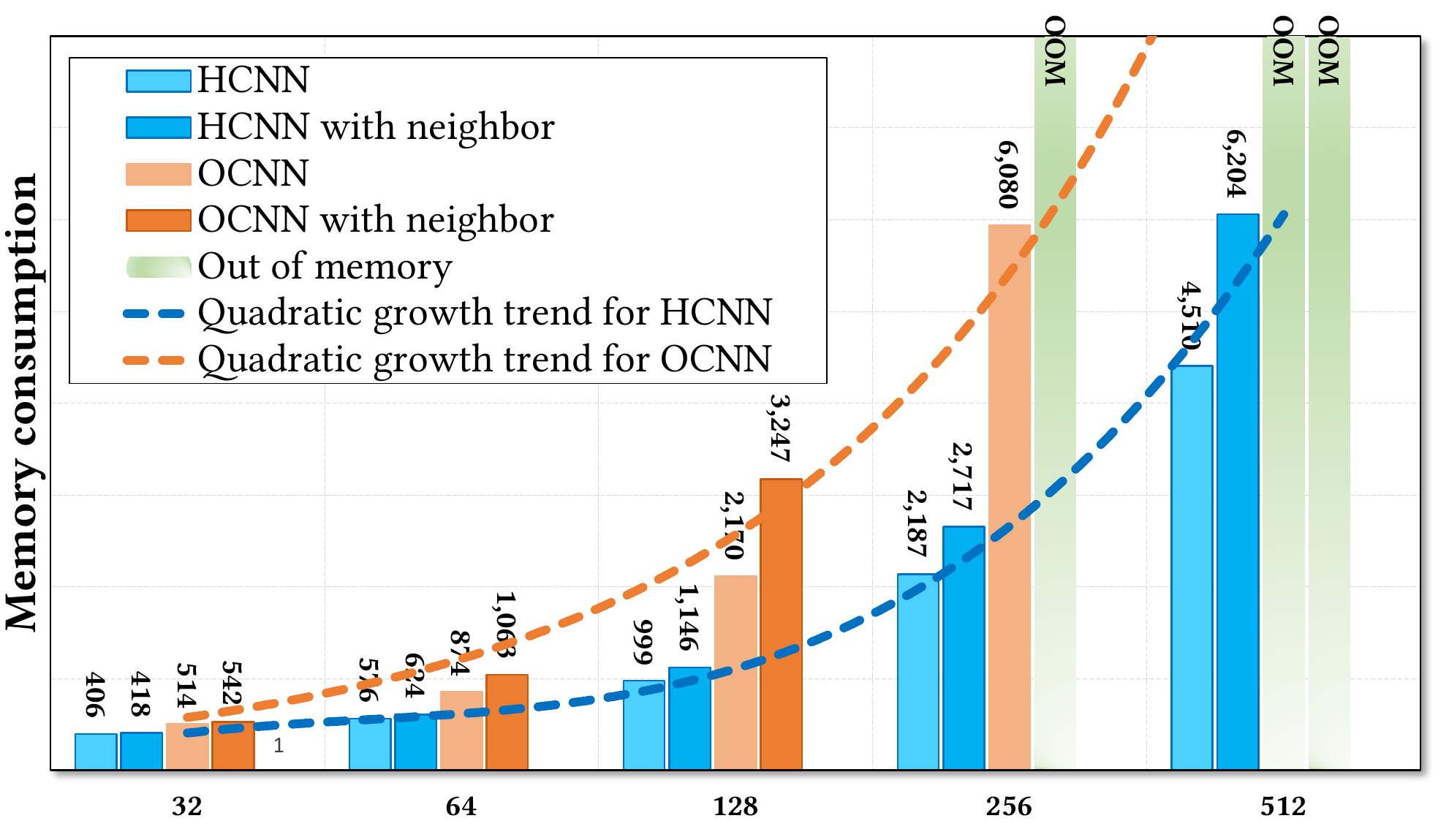}
  \caption{The actual memory consumption using OCNN and HCNN over a mini-batch of $32$ models. The physical memory cap of the $\mathtt{1080}$ $\mathtt{GTX}$ card is $8$ GB. HCNN allows very dense voxelization up to $512^3$ even with pre-stored neighbor information, while OCNN can only handle resolution of $128^3$ with recorded neighborhood.}\label{fig:memory_minibatch}
\end{figure}

In reality, the memory footprint follows the similar pattern. \figref{fig:memory_minibatch} compares the memory usage for OCNN and HCNN during the mini-batch training. A mini-batch consists of $32$ random models, and memory usage is quite different for different batches. Therefore, we report the batch which uses the largest amount of memory during 1,000 forward and backward iterations. It can be seen from the figure that when the resolution is $256^3$, OCNN consumes $6,080$ MB memory, and our method just needs $2,187$ MB memory. This is over $170\%$ less memory consumption. When the resolution is further increased to $512^3$, OCNN is unable to fit the entire batch into 8 GB  memory of the \texttt{1080 GTX} video card, while our method is not even close to the cap, which only uses $4,510$ MB memory. If one chooses to use the entire voxel grid, a mini-batch would need over 2 GB memory (with \texttt{nVidia cuDNN}) under resolution of $64^3$, which is roughly four times of HCNN. During CNN training, one could accelerate the convolution-like operations by saving the neighborhood information for each non-empty voxel (or each leaf-octant with OCNN). With this option enabled, OCNN is even not able to handle the batch under $128^3$, while our method is sill able to deal with the batch under $512^3$. The plotted growth trends also suggest that the gap of the memory consumption between OCNN and HCNN should be quickly widened with the increased voxel resolution.

\begin{table}[ht!]
\begin{center}
\begin{tabular}{l|c|c}
    \whline{1.15pt}
\textbf{Network architecture} & \textbf{Without voting} & \textbf{With voting } \\
 \whline{0.65pt}
HCNN(32) & \textcolor{blue}{$89.3\%$} & $89.6\%$   \\
OCNN(32) & \textcolor{blue}{$89.3\%$} & \textcolor{blue}{$89.8\%$}   \\
FullVox(32) & \textcolor{blue}{$89.3\%$} & \textcolor{blue}{$89.8\%$} \\
\hline
HCNN(64) & \textcolor{blue}{$89.3\%$} & \textcolor{blue}{$89.9\%$}   \\
OCNN(64) & \textcolor{blue}{$89.3\%$} & $89.8\%$  \\
FullVox(64) & $89.0\%$ & $89.6\%$ \\
\hline
HCNN(128) & \textcolor{blue}{$89.4\%$} & \textcolor{blue}{$90.1\%$} \\
OCNN(128) & $89.2\%$ & $90.0\%$  \\
\hline
HCNN(256) & \textcolor{blue}{$89.2\%$} & \textcolor{blue}{$90.2\%$}  \\
OCNN(256) & \textcolor{blue}{$89.2\%$} & \textcolor{blue}{$90.2\%$}  \\
\hline
HCNN(512) & $89.1\%$ & $89.6\%$  \\
OCNN(512) & OOM & OOM  \\
\hline\hline
VoxNet(32) & $82.0\%$ & $83.0\%$ \\
Geometry image & $83.9\%$ &  -- \\
SubVolSup(32) & $87.2\%$ &  $89.2\%$ \\
FPNN(64) & $87.5\%$ &  -- \\
PointNet & $89.2\%$ &  -- \\
VRN(32) & $89.0\%$ & \textcolor{blue}{$91.3\%$}\\

\whline{1.15pt}
\end{tabular}
\end{center}
\caption{Benchmark of shape classification on \texttt{ModelNet40} dataset. In the forst portion of the table, we report the classification results using HCNN and OCNN. The classification accuracy using fully voxelized models (FullVox) is also reported. The number followed by a network name indicates the resolution of the discretization. In the second half of the table, the benchmarks of other popular nets are listed for the comparison. The best benchmark among a given group is highlighted in \textcolor{blue}{blue} color.}\label{tab:classification}
\end{table}

\subsection{Shape Classification}\label{subsec:classification}
The first shape analysis task is the shape classification, which returns a label out of a pre-defined list that best describes the input model. The dataset used is \texttt{ModeNet40}~\cite{wu20153d} consisting of 9,843 training models and 2,468 test models. The upright direction for each model is known, and we rotate each model along the upright direction uniformly generating 12 poses for the training. At the test stage, the scores of these 12 poses can be pooled together to increase the accuracy of the prediction. This strategy is known as \emph{orientation voting}~\cite{maturana2015voxnet}. The classification benchmarks of HCNN under resolutions from $32^3$ to $512^3$ with and without voting are reported in~\tabref{tab:classification}.

In the first half of the table, we also list the prediction accuracy using OCNN~\cite{OCNN_2017} and FullVox under the same resolution. The notion of HCNN(32) in the table means the highest resolution of the HCNN architecture is $32^3$. The so-called FullVox refers to treating a 3D model as a fully voxelized bounding box, where a voxel either houses the corresponding normal vector, as HCNN or OCNN does, if it intersects with the model's surface, or a zero vector. In theory, FullVox explicitly presents the original geometry of the model without missing any information -- even for empty voxels. All the CNN operations like convolution and pooling are applied to both empty and non-empty voxels. This na\"ive discretization is not scalable and becomes prohibitive when the voxel resolution goes above $64^3$. The reported performance of OCNN is based on the published executable at~\url{https://github.com/Microsoft/O-CNN}. As mentioned above, we shape our HCNN architecture to exactly match the one used in OCNN to avoid any influences brought by different networks. We can see from the benchmarks that under moderate resolutions like $32^3$ and $64^3$, HCNN, OCNN and FullVox perform equally well, and employing the voting strategy is able to improve the accuracy by another five percentages on average. When the voxel resolution is further increased, overfitting may occur as pointed out in~\cite{OCNN_2017}, since there are no sufficient training data to allow us to fine-tune the network's parameters. As a result, the prediction accuracy slightly drops even with voting enabled.

The second half of \tabref{tab:classification} lists the classification accuracy of some other well-known techniques including VoxNet~\cite{maturana2015voxnet}, Geometry image~\cite{sinha2016deep}, SubVolSup~\cite{qi2016volumetric}, FPNN~\cite{li2016fpnn}, PointNet~\cite{pointnet} and VRN~\cite{brock2016generative}. We also noticed that the performance of OCNN in our experiment is slightly different from the one reported in the original OCNN paper. We suspect that this is because different parameters used during the model rectification stage (i.e. the magnitude of the dilation).

\begin{table}[ht!]
\begin{center}
\begin{tabular}{l|c|c|c|c|c}
    \whline{1.15pt}
\textbf{Network architecture} & $32^3$ & $64^3$ & $128^3$ & $256^3$ & $512^3$ \\
 \whline{0.65pt}
HCNN & \textcolor{blue}{25.2} & \textcolor{blue}{73.1} & \textcolor{blue}{217.3} & \textcolor{blue}{794.3} & \textcolor{blue}{2594.2}
\\
OCNN & 27.5 & 78.8 & 255.0 & 845.3 & OOM \\
\hline
OCNN with neighbor & 24.0 & 72.0 & 244.4 & OOM & OOM \\
HCNN with neighbor & \textcolor{blue}{22.9} & \textcolor{blue}{67.9} & \textcolor{blue}{205.4} & \textcolor{blue}{772.7} & \textcolor{blue}{2555.5} \\
\hline
FullVox & 39.7 & 269.0 & OOM & OOM & OOM \\

\whline{1.15pt}
\end{tabular}
\end{center}
\caption{Average forward-backward iteration speed using HCNN, OCNN and FullVox (in $ms$). For a fair comparison, we exclude the hard drive I/O time.}\label{tab:time}
\end{table}

Compact hashing also improves the time performance of the networks. \tabref{tab:time} reports the average forward-backward time in $ms$ over $1,000$ iterations. We can see that HCNN is consistently faster than OCNN regardless if the neighbor information is pre-recorded, not to mention the FullVox scheme. The timing information reported does not include the hard drive I/O latency for a fair comparison. In our experiment, HCNN is typically $10\%$ faster than OCNN under the same resolution.

\subsection{Shape Retrieval}\label{sec:retrieval}
The next shape analysis task is the shape retrieval.
In this experiment, we use the \texttt{ShapeNet} \texttt{Core55} dataset, which consists of $51,190$ models with $55$ categories. Subcategory information associated with models is ignored in this test. $70\%$ of the data is used for training; $10\%$ is used for validation, and the rest $20\%$ is for testing. Data augmentation is performed in this test by rotating 12 poses along the upright direction for each model. The orientational pool is also used~\cite{qi2016volumetric,su2015multi}. The neural network produces a vector of the category probability scores for an input query model, and the model is considered belonging to the category of the highest score. The retrieval set corresponding to this input query shape is a collection of models that have the same category label sorted according to the L-2 distance between their feature vectors and the query shape's. Precision and recall are two widely-used metrics, where precision refers to the percentage of retrieved shapes that correctly match the category label of the query shape, and recall is defined as the percentage of the shapes of the query category that have been retrieved. For a given query shape, with more instances being retrieved, the precision drops when a miss-labeled instance is retrieved. On the other hand, recall quickly goes up since more models out of the query category have been retrieved.
\begin{figure}[t!]
  \centering
  \includegraphics[width=\linewidth]{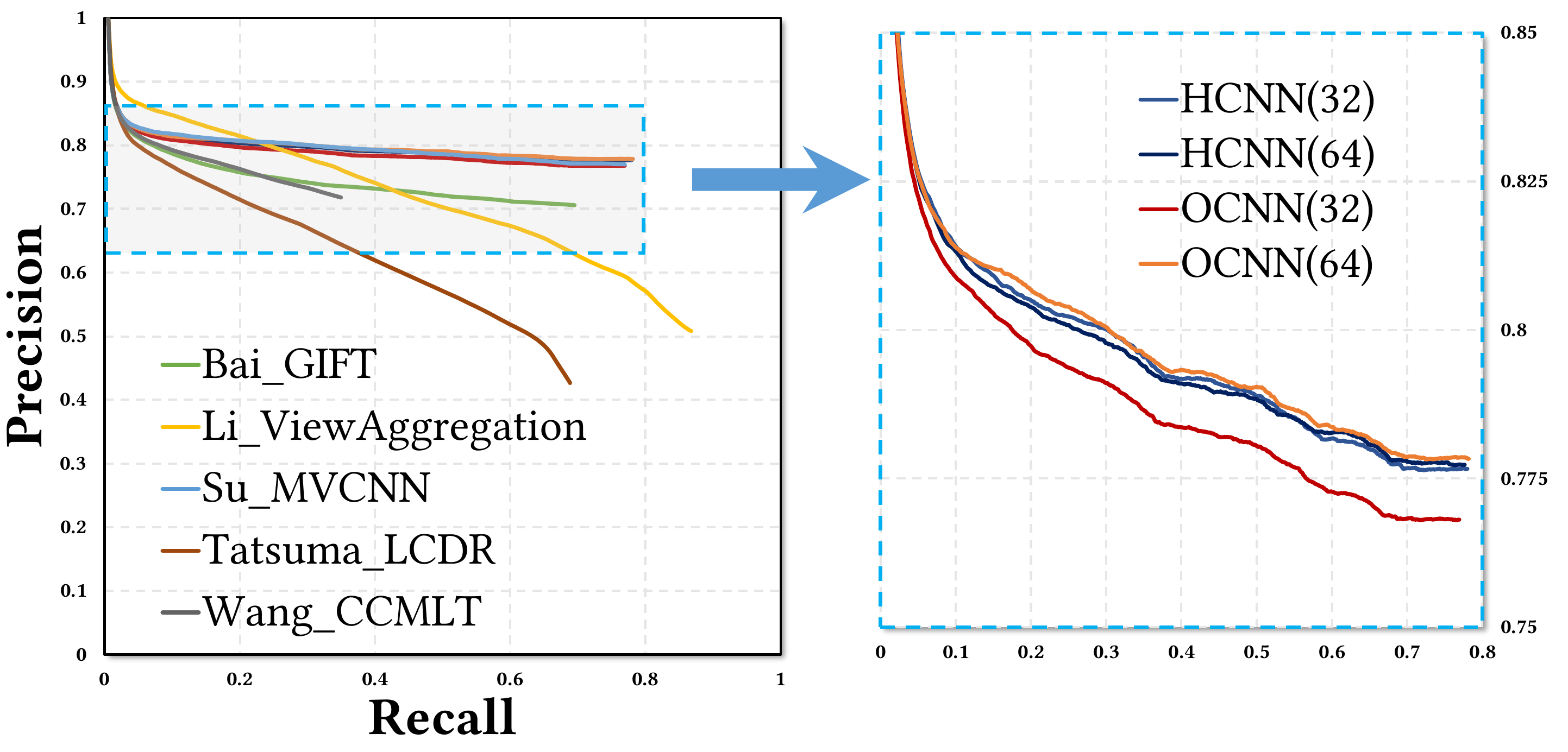}
  \caption{The precision recall curves for HCNN, OCNN as well as other five famous multi-view CNN methods from SHREC16. The difference between HCNN and OCNN (under resolutions of $32^3$ and $64^3$) is quite subtle even after zooming in.
  }\label{fig:pr_curve}
\end{figure}

The comparative precision and recall curves are shown in \figref{fig:pr_curve}. Together with our HCNN under resolutions of $32^3$ and $64^3$, we also plot the curves for OCNN(32) and OCNN(64) as well as several widely-known methods including GIFT~\cite{bai2016gift}, Multi-view CNN~\cite{su2015multi}, Appearance-based feature
extraction using pre-trained CNN and Channel-wise CNN~\cite{savva2016shrec}. The performance benchmarks of these latter methods are obtained using the published evaluator at \url{https://shapenet.cs.stanford.edu/shrec16/}. From the figure, we can see that 3D CNN methods like HCNN and OCNN outperform multi-view based methods, since the geometry information of the original models is much better encoded. The performances of HCNN and OCNN are very close to each other. After enlarging curve segments associated with HCNN(32), HCNN(64), OCNN(32) and OCNN(64) within the precision interval of $[0.75, 0.85]$, one can see that OCNN(32) is slightly below (worse) the other three.

\setlength{\columnsep}{5 pt}
\begin{wrapfigure}{r}{0.5\linewidth}
\includegraphics[width =\linewidth]{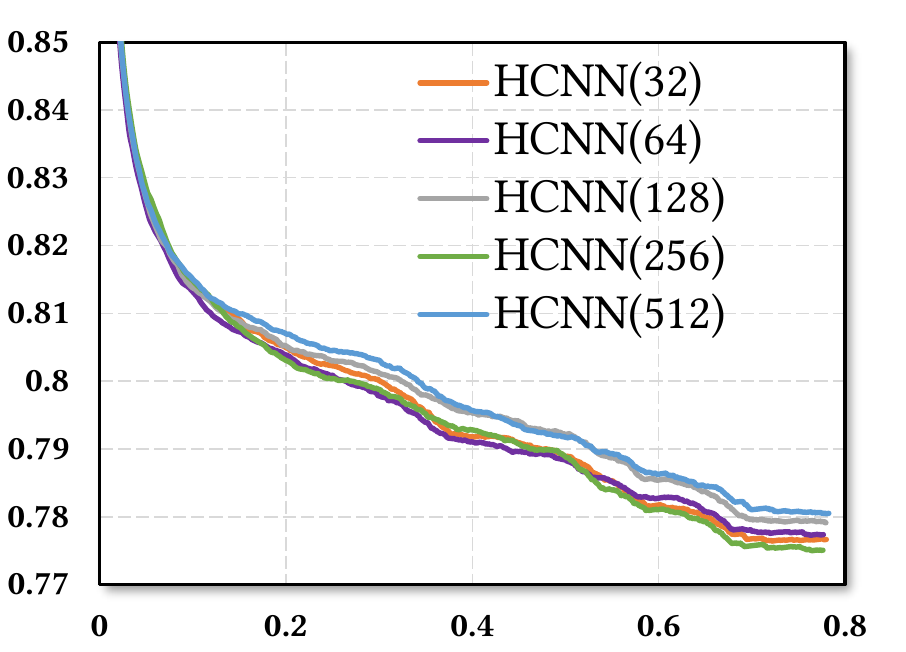}
\vspace{-15 pt}
\end{wrapfigure}
Another interesting finding is that HCNN seems to be quite inert towards the voxel resolution. As shown on the right, HCNN(32) already has a very good result while further increasing the resolution to $512^3$ does not significantly improve the performance. Curves for HCNN(32) to HCNN(512) are hardly discernible. We feel like this actually is reasonable since identifying a high-level semantic label of an input 3D model does not require detailed local geometry information in general -- even a rough shape contour may suffice. Similar conclusion can be drawn when evaluating the retrieval performance using other metrics as reported in \tabref{tab:retrieval}. Here in addition to precision and recall, we also compare the retrieval performance in terms of mAP, F-score and NDCG, where mAP is the mean average precision, and F-score is the harmonics mean of the precision and recall. NDCG reflects the ranking quality and the subcategory similarity. It can be seen from the table that, HCNN has a comparable performance as OCNN does. Both outperform multi-view based methods.

\begin{table}[ht!]
\begin{center}
\begin{tabular}{l|c|c|c|c|c}
    \whline{1.15pt}
\textbf{Method} & \textbf{P@N} & \textbf{R@N} & \textbf{mAP} & \textbf{F1@N} & \textbf{NDCG}  \\
 \whline{0.65pt}
Tatsuma\_LCDR & 0.427 & 0.689 & 0.728 & 0.472 & 0.875\\
Wang\_CCMLT & 0.718 & 0.350 & 0.823 & 0.391 & 0.886   \\
Li\_ViewAgg. & 0.508 & \textcolor{blue}{0.868} & 0.829 & 0.582 & 0.904  \\
Bai\_GIFT & 0.706 & 0.695 & 0.825 & 0.689 & 0.896 \\
Su\_MVCNN & 0.770 & 0.770 & 0.873 & 0.764 & 0.899 \\
\hline\hline
OCNN(32) & 0.768 & 0.769 & 0.871 & 0.763 & 0.904 \\
OCNN(64) & 0.778 & 0.782 & 0.875 & 0.775 & 0.905 \\
\hline
HCNN(32) & 0.777 & 0.780 & 0.877 & 0.773 & 0.905 \\
HCNN(64) & 0.777 & 0.777 & \textcolor{blue}{0.878} & 0.772 & 0.905 \\
HCNN(128) & 0.778 & 0.779 & \textcolor{blue}{0.878} & 0.774 & \textcolor{blue}{0.906} \\
HCNN(256) & 0.775 & 0.776 & \textcolor{blue}{0.878} & 0.773 & \textcolor{blue}{0.906} \\
HCNN(512) & \textcolor{blue}{0.780} & 0.783 & 0.874 & \textcolor{blue}{0.777} & \textcolor{blue}{0.906} \\

\whline{1.15pt}
\end{tabular}
\end{center}
\caption{Shape retrieval benchmarks.}\label{tab:retrieval}
\end{table}
\begin{figure*}[t!]
  \centering
  \includegraphics[width=\linewidth]{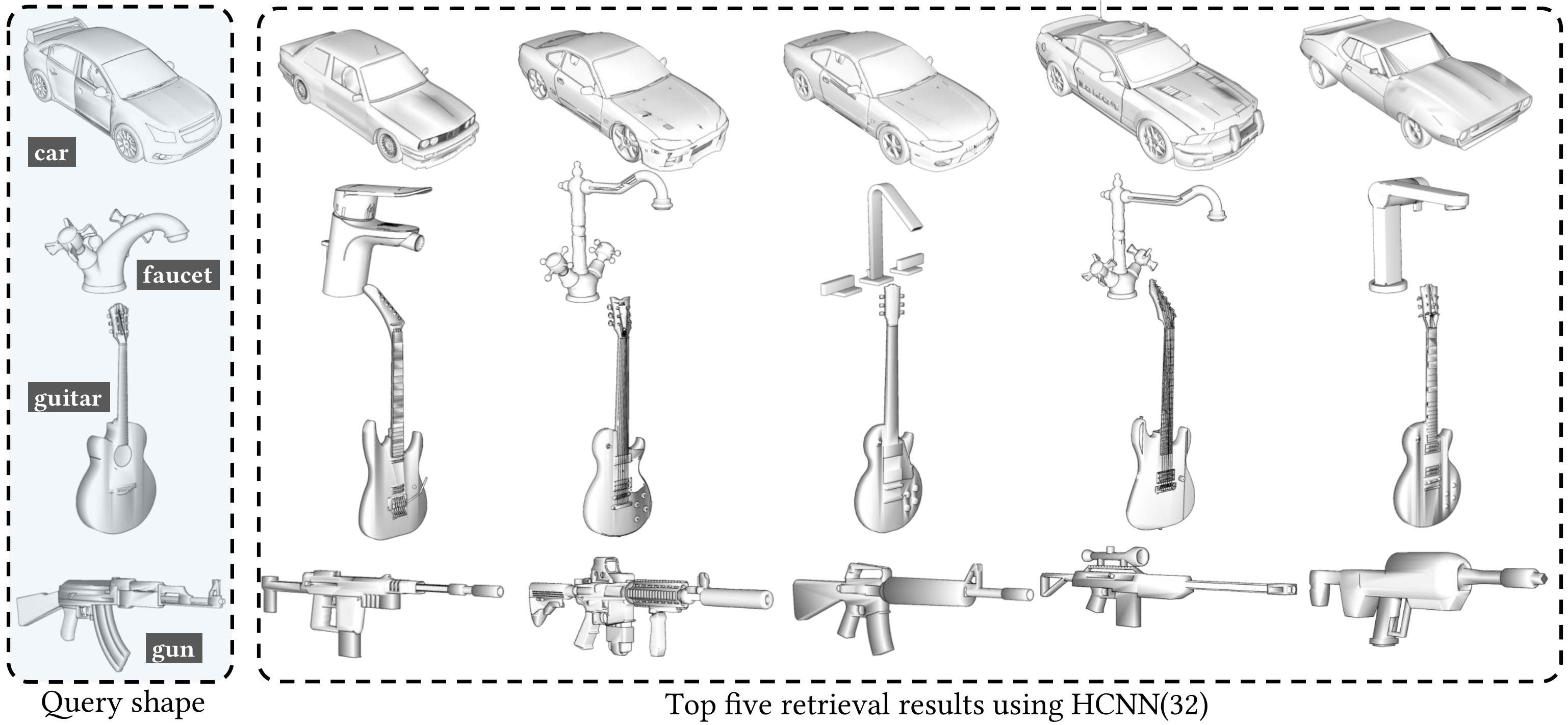}
  \caption{The top five retrieval result of input queries of four categories, namely car, faucet, guitar and gun. The leftmost column is the query input.
  }\label{fig:retrieval}
\end{figure*}

\subsection{Shape segmentation}
\begin{figure}[t!]
  \centering
  \includegraphics[width=\linewidth]{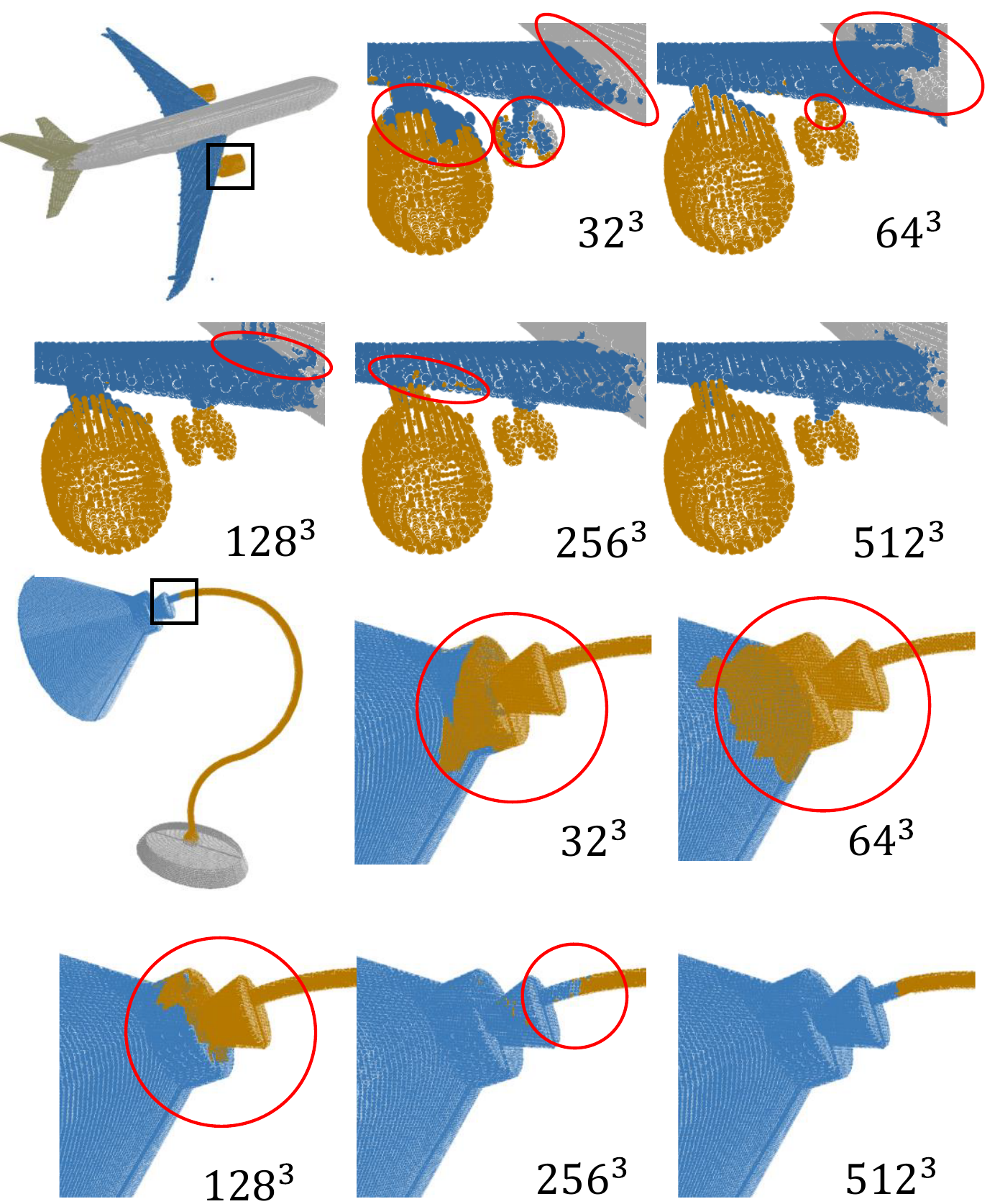}
  \caption{In shape segmentation, high voxel resolution better captures local geometry features between adjacent object parts and yields better results.
  }\label{fig:segmentation}
\end{figure}

Finally, we discuss the experimental results of the shape segmentation, which assigns each point or triangle on the input model a part category label. Our experiment is based on the dataset in~\cite{yi2016scalable}, which adds extra semantic part annotations over a subset of models from ShapeNet. The original dataset includes 16 categories of shapes, and each category has two to six parts. Clearly, segmentation is more challenging than classification or retrieval since part segmentation often relies on local geometry features, and we would like to fully test the advantage of the high-resolution voxelization that is only possible with HCNN. On the other hand, more hierarchy levels also induce more network parameters to be tuned during the CNN training. Therefore, we only test the segmentation performance when there are sufficient training data. Again, we rotate 12 poses along the upright direction for each model to augment the dataset. The training/test split is set the same as in~\cite{OCNN_2017}. We consider the segmentation as a per-point classification problem, and use intersection over union (IoU) to quantitatively evaluate the segmentation quality as did in~\cite{pointnet}.
It is noteworthy that the statistics reported in~\cite{OCNN_2017} were actually based on IoU counts on the leaf octants. It is easy to understand that under moderate voxel resolutions, the mean IoU (mIoU) defined on the voxel grid trends to have a better benchmark than on the original point cloud because a coarse discretization could alias the true boundary between two segments on the original model. To avoid such confusion, we report benchmarks of HCNN under different resolutions on both voxel grids and the input point clouds (i.e. the so-called HCNNP in the table) in \tabref{tab:segmentation}. We can see that discretizing models at higher resolutions effectively improves the segmentation quality. While the mIoU improvement may read incremental from the table, those improvements lead to a better classification of points near the segmentation boundary. As shown in \figref{fig:segmentation}, the segmentation result improvement is visually noticeable with higher voxel resolutions.

\begin{table}[ht!]
\begin{center}
\begin{tabular}{l|c|c|c|c|c|c}
    \whline{1.15pt}
\textbf{Method} & \textbf{Plane} & \textbf{Car} & \textbf{Chair} & \textbf{Guitar} & \textbf{Lamp} & \textbf{Table}\\
 \whline{0.65pt}
Yi et al. 2016 & $81.0\%$ & $75.7\%$ & $87.6\%$ & $92.0\%$ & $82.5\%$ & $75.3\%$\\
PointNet & $83.4\%$ & $74.9\%$ & $89.6\%$ & $91.5\%$ & $80.8\%$ & $80.6\%$\\
SpecCNN & $81.6\%$ & $75.2\%$ & $90.2\%$ & \textcolor{blue}{$93.0\%$} & \textcolor{blue}{$84.7\%$} & $82.1\%$\\
\hline
OCNN(32) & $84.2\%$ & $74.1\%$ & $90.8\%$ & $91.3\%$ & $82.5\%$ & $84.2\%$\\
OCNN(64) & $85.5\%$ & $77.0\%$ & $91.1\%$ & $91.9\%$ & $83.3\%$ & $84.4\%$\\
\hline
HCNN(32) & $85.4\%$ & $75.8\%$ & $91.3\% $ & $91.8\%$ & $83.3\%$ & $85.8\%$\\
HCNN(64) & $85.5\%$ & $77.0\%$ & $91.3\% $ & $92.0\%$ & $83.7\%$ & $85.7\%$\\
HCNN(128) & $85.6\%$ & $78.7\%$ & $91.3\%$ & $92.0\%$ & $83.6\%$ & $85.9\%$\\
HCNN(256) & $85.8\%$ & $79.3\%$ & \textcolor{blue}{$91.4\%$} & $92.0\%$ & $84.0\%$ & \textcolor{blue}{$86.0\%$}\\
HCNN(512) & \textcolor{blue}{$86.8\%$} & \textcolor{blue}{$80.2\%$} &  $91.3\%$ & $91.9\%$ & $84.0\%$ & $85.9\%$\\
\hline
HCNNP(32) & $81.1\%$ & $77.2\%$ & $90.7\%$ & $90.8\%$ & $83.2\%$ & $85.3\%$\\
HCNNP(64) & $85.0\%$ & $78.9\%$ & $91.5\%$ & $91.7\%$ & $83.8\%$ & $85.9\%$\\
HCNNP(128) & $86.2\%$ & $79.9\%$ & \textcolor{blue}{$91.8\%$} & $91.9\%$ & $83.9\%$ & $86.2\%$ \\
HCNNP(256) & $86.3\%$ & $79.8\%$ & \textcolor{blue}{$91.8\%$} & $92.0\%$ & $84.1\%$ & $86.1\%$ \\
HCNNP(512) & \textcolor{blue}{$86.9\%$} & \textcolor{blue}{$80.1\%$} & \textcolor{blue}{$91.8\%$} &  $91.9\%$ & $84.3\%$ & \textcolor{blue}{$86.2\%$}\\

\whline{1.15pt}
\end{tabular}
\end{center}
\caption{Benchmarks for shape segmentation. HCNNP($\cdot$) refers to the benchmarks based on IoU counts over the original input point clouds under the corresponding voxel resolution.}\label{tab:segmentation}
\end{table}

\vspace{5 pt}
\noindent\textbf{Discussion}\hspace{5 pt}
In summary, as clearly demonstrated in our experiments, HCNN acts like a ``superset'' of OCNN, which we consider as the most state-of-the-art 3D CNN method and our closest competitor. The benchmarks in different shape analysis tasks of using our HCNN are at least very comparable to the ones obtained using OCNN, if not better. However, we would like to remind the reader that the memory consumption of HCNN is significantly less than OCNN and the time performance is also slightly better (i.e. $\sim10\%$ as reported in \tabref{tab:time}). As a result, HCNN allows 3D CNNs to take high-resolution models during the training. For shape classification and retrieval, the primary task for the neural network is to \emph{reduce} a complex input 3D model to few semantic labels i.e. from a very high-dimension vector to a low-dimension one. It is not surprising to us that a dense voxelization has limited contributions towards the final benchmark. On the other hand, a high-quality segmentation requires detailed local geometry features which is somewhat commensurate to the voxel resolution. Therefore, increasing the resolution improves segmentation result in general. Undoubtedly, being able to input high-resolution models to the CNN will broaden the 3D CNN applications and potentially allow us to leverage CNNs to deal with more challenging tasks for 3D graphics contents such as shape synthesis~\cite{fisher2012example,ying2001texture}. Besides what has been discussed in the experiment, our HCNN is more versatile and is compatible with all CNN configurations like arbitrarily-strided convolution and pooling. While not yet particularly popular, research efforts have already been devoted to investigate the advantages of such irregular CNNs~\cite{han2017optimizing}. Our HCNN would facilitate such possible future research endeavors more friendly.

\section{Conclusion}
\label{sec:conclusion}
In this paper, we present a novel 3D CNN framework, named HCNN, for high-resolution shape analysis. Our data structure constructs a set of hierarchical perfect spatial hashing of an input 3D model at differen resolutions. HCNN is memory-efficient, and its memory overhead is polynomially smaller than existing octree-based methods like OCNN~\cite{OCNN_2017}. We test the proposed HCNN for three classic shape analysis tasks: classification, retrieval and segmentation. The experimental results show that HCNN yields similar or better benchmarks compared with state-of-the-art, while reducing the memory consumption up to three times.

Currently, all the PSHs are generated using the CPU. In the future, we would like to use the GPU to further accelerate this procedure~\cite{alcantara2009real}. Thanks to its superior memory performance, HCNN allows high-resolution shape analysis that is not possible with existing methods, which paves the path of using CNN to deal with more challenging shape analysis tasks such as shape synthesis, denoising and morphing. Since HCNN allows irregular CNN configurations, we are also interested in optimizing the network structure~\cite{lorenzo2017hyper} by fine-tuning hyper-parameters.


\ifCLASSOPTIONcaptionsoff
  \newpage
\fi

\bibliographystyle{IEEEtran}
\bibliography{HCNN}


%



\end{document}